# Confocal microphotoluminescence mapping of coupled and detuned states in photonic molecules

F.S.F. Brossard,[1]* B.P.L. Reid,[2] C.C.S. Chan,[2] X. L. Xu,[1] J. P. Griffiths,[3] D.A. Williams,[1] R. Murray,[4] and R.A. Taylor[2]

[1] Hitachi Cambridge Laboratory, Hitachi Europe Ltd., Cambridge CB3 0HE, UK

[2] Department of Physics, University of Oxford, Oxford OX1 3PU, UK

[3] Cavendish Laboratory, University of Cambridge, Cambridge, CB3 0HE, UK

[4] Department of Physics, Imperial College, London SW7 2AZ, UK

## Abstract

**Abstract:** We study the coupling of cavities defined by the local modulation of the waveguide width using confocal photoluminescence microscopy. We are able to spatially map the profile of the antisymmetric (antibonding) and symmetric (bonding) modes of a pair of strongly coupled cavities (photonic molecule) and follow the coupled cavity system from the strong coupling to the weak coupling regime in the presence of structural disorder. The effect of disorder on this photonic molecule is also investigated numerically with a finite-difference time-domain method and a semi-analytical approach, which enables us to quantify the light localization observed in either cavity as a function of detuning.

Photonic molecules (PM) are essentially two evanescently coupled optical cavities and are so-called due to the similarity between their coupled states and that found in diatomic molecules [1]. The properties of these PM have led to applications particularly relevant for on-chip devices such as bistable [2] or switchable [3] lasers, channel drop filters [4-5] and optical memories [6] amongst others. Recently, they have also found applications in quantum optics such as the production of entangled photon pairs when coupled to quantum dots [7] or of single photons with the aid of a relatively weak non-linearity [8-10]. Photonic crystal (PhC) cavities are particularly suitable in this respect due to the small mode volume and high quality factor found in these engineered defects [11]. When embedded in a PhC waveguide (PhCWG) they can form large-scale array of ultra-high Q coupled resonators [12] offering the possibility of studying quantum many-body phenomena [13,14]. However, the coupling between individual cavities can be highly susceptible to manufacturing imperfections [15]which can strongly modify the intensity distribution within the molecule, as demonstrated with the D2 [16], L3 [17-20] and nanobeam PM [21]. The consequences are far-reaching when considering light-matter coupling with a quantum emitter as this critically depends on the local strength of the cavity fields [22]. It is therefore crucial to clarify the effect of such imperfect coupling on the mode distribution within the studied PM for applications in cavity quantum electrodynamics (cQED).

Previous work on coupled PhCWG cavities was motivated by the engineering of optical delay lines using coupled resonator optical waveguides (CROWs) which involved analysing the transmitted signal through the PhC waveguide to demonstrate slow light [12,15]. In this article our work is primarily concerned with providing information about the field distribution within coupled PhCWG cavities for applications in light-matter coupling and assessing the tolerance of this type of coupled system to structural disorder. In order to achieve this we use a confocal photoluminescence microscopy mapping technique to first demonstrate clearly two strongly coupled PhCWG cavities from the symmetry of the supermodes. Next, we investigate experimentally the effect of detuning on the PM mode distribution, and use this information to model numerically both the detuning and the degree of coupling. Our work complements that in [21] on nanobeam cavities by quantifying the effect of detuning on the delocalization, by measuring the intensity found in each PhCWG cavity for a particular supermode as a function of detuning. This enables us to follow the coupled cavity system from the strong coupling to the weak coupling regime. Throughout this article experimental results are compared with numerical calculations obtained from a freely available three-dimensional (3D) finite-difference time-domain (FDTD) package MEEP [23] using a filter diagonalization method to extract the resonances [24].

The PM studied here consists of two adjacent in-line cavities separated by 6 lattice units centre-to-centre between the cavities and is based on the local modulation of the width of a PhCWG with a triangular lattice of air holes [25]. Each cavity is formed by gradually shifting the holes away from the PhCWG as shown in Fig. 1(a), effectively creating a local optical well with ultra-low out-of-plane losses in a manner similar to a double heterostructure [26]. The devices were fabricated in 200 nm thick GaAs air membrane obtained by the removal of a 1.3 μm thick $Al_{0.7}Ga_{0.3}As$ sacrificial layer with HF treatment. The air holes were obtained using a lithographic step with an ultra-high resolution 100 kV VB6 Leica e-beam lithography system followed by reactive ion etching (RIE). Details of the fabrication can be found in our previous work on this cavity [27].

Simulations were performed on the PMs using structural parameters extracted from scanning electron microscope (SEM) images of fabricated devices. The results shown in Figs. 1(b) and 1(c) indicate the existence of a symmetric (S) and an antisymmetric (AS) supermode resulting from the linear superposition of the fundamental mode of two isolated cavities (shown superimposed for comparison). This is expected from the physics governing the coupling between harmonic oscillators where the S/AS supermode is the result of in-phase/out-of-phase oscillations between each cavity [1]. Importantly the S and AS supermodes are clearly distinguishable here due to the presence of a large antinode for the S supermode at the centre of the molecule. Such an antinode is complicated by field discontinuities in coupled PhC cavities separated by hole-barriers [18]. In these devices, the S and AS supermodes can be distinguished by the phase sensitivity of far field measurement techniques [19]. In addition, the single cavity version of the device studied in this work shown in Fig. 1(d) with its first order mode shown in Fig. 1(e) supports a higher order bound mode of odd parity, whose mode distribution closely matches that of the AS supermode as shown in Fig. 1(f). Measurements from a single cavity are presented to remove ambiguities as to the origin of the modes observed in the PM.

The fabricated structures were investigated using a confocal microphotoluminescence (μPL) mapping technique on a high InAs dot density (~200 dots/μm$^2$) wafer with broad emission centred at 1200 nm and spectral bandwidth ~ 160 nm. The samples were cooled to 5 K within a continuous-flow liquid helium cryostat and the dots used as an internal light source excited by a He-Ne laser (632.8 nm) with estimated power of 50 μW at the sample surface. For the mapping, a 100× microscope objective (numerical aperture 0.5) with 12 mm working distance giving a ~ 1.2 μm spot size was mounted on a piezoelectric *xyz* stage with 1 nm resolution and data acquired in 200 nm steps along the PhCWG direction. The emitted light was collected by the same objective, dispersed through a 0.3 m spectrometer and detected by a cooled 512 pixel InGaAs array. This setup was previously described in detail and used to register the position of a single quantum dot with high accuracy [28]. Fig. 2(a) shows the footprint of a fabricated PhCWG presented in Fig. 2(b) as determined by the PL mapping technique. The scanning direction of the piezo stage was modified with a computer program to closely match that of the submicron wide waveguide section of the PhC clearly visible in Fig. 2(a). This guaranteed scanning through the center of each cavity with a few line scans only along the axis of the waveguide. The results presented hereafter are from the line scan giving maximum signal intensity from the cavity resonances.

The intensity emission profiles of the first and second order bound modes of a single cavity are shown in Fig. 3(a). The computed profiles given in Fig. 3(b) were obtained by convolving a Gaussian beam of 1.2 μm size with the time averaged electric field energy distribution given by the 3D FDTD method for comparison. The wavelengths of the resonance peaks were computed applying the filter diagonalization method to the FDTD calculated E field time traces. They were also confirmed by a separate calculation involving transmission through the device, in both cases using $n = 3.33$ for the index of GaAs [29]. These results show that we can resolve the two maxima of the higher energy mode located at the edge of the cavity, the signature of the second order mode. In contrast, the fundamental mode has a single maximum at the centre of the cavity. A spectral separation of 8.6 nm between the first and second order modes was measured from the PL spectra in Fig. 3(a), in excellent agreement with the theoretical value of 8.0 nm as shown in Fig. 3(b).

Fig. 3(c) shows the intensity emission profile of the modes distributed over a PM. In this case, we observe two modes extending over both cavities: a low energy mode with a single maximum localized near the centre of the coupled cavity system and a higher energy mode spectrally separated by 1.3 nm with a maximum in each cavity. The two maxima are separated by 2 μm as expected from the centre-to-centre distance between the cavities. These results are in very good agreement with the computed mode profile of the first order S and AS supermodes of a PM shown in Fig. 3(d) and calculated spectral splitting of 1.8 nm, thus indicating a strongly coupled system due to coupling between the first order mode of individual cavities. This splitting is much smaller than the spectral separation between the first and second order mode of a single cavity. This information enables us to distinguish the AS supermode from the second order mode, without which a weakly coupled system could be mistaken for a strongly coupled one. We also note that the second order mode produces two additional S and AS supermodes (not shown here) when the cavities are strongly coupled. However, these supermodes never crossed the wavelength range of the lower order mode (coupled or not), so that no confusion was possible. We also observe that the two maxima seen in the intensity distribution of the S supermode merge into a single peak because of convolution with the resolution function of the collection lens. The latter effect enables us to clearly distinguish the S from the AS supermode experimentally in this PM without requiring information about the phase property of these modes.

Only about a third of the nominally identical PMs investigated showed evidence of a strongly coupled system. Indeed it is well known that the coupling between PhC cavities can be greatly affected by fabrication imperfections resulting in mode localization [16-21]. In Fig. 4(a) we present the spectral splitting measured from a batch of devices and the typical intensity emission profile of selected coupled cavities in Figs. 4(b)-4(d), showing increasingly localized modes in each cavity. Evidence of a delocalized mode was generally observed for a smaller splitting of ~ 1 nm. Coupled cavities with a splitting of ~ 2 nm showed some degree of mode localization whilst giving a distinguishable intensity emission profile between the high and low order modes. Devices with a splitting of ~ 3 nm systematically showed clear mode localization with little or no distinction between the profiles found in either cavity. The latter case gave modes with a similar profile to those found in isolated cavities.

To understand this behaviour we recall some basic theory of coupled oscillators considering the coupled-cavity as a two-level system [30]. When cavities are brought together, the coupling generates a shift in their resonant wavelength with a splitting given by (neglecting losses):

$$\Delta\Omega = \sqrt{(\Delta\lambda)^2 + 4J^2}$$
$$\Delta\lambda = \lambda_1 - \lambda_2$$

(1)

where $\lambda_1$ and $\lambda_2$ are the resonant wavelengths of isolated cavities, $\Delta\lambda$ is the detuning between the cavities and $J$ is the coupling strength due to mode overlap. Hence one can deduce the coupling strength in the case of zero detuning and subsequently deduce the detuning from the splitting provided $J$ is conserved. The modes of the coupled system are obtained from the linear combination of the modes in isolated cavities as defined by:

$$\psi_+(r) = \left(\sin\frac{\theta}{2}\right)\varphi_2(r) + \left(\cos\frac{\theta}{2}\right)\varphi_{1(r)}$$

$$\psi_-(r) = \left(\cos\frac{\theta}{2}\right)\varphi_2(r) - \left(\sin\frac{\theta}{2}\right)\varphi_{1(r)} \qquad (2)$$

$$\tan\theta = \frac{2J}{\Delta\lambda}$$

where $\varphi_{1(r)}$ and $\varphi_{2(r)}$ refer to the wavefunctions of the modes in isolated cavities, $\psi_{+(r)}$ and $\psi_{-(r)}$ the wavefunctions of the coupled system. The weighting factor of each mode $\varphi_{1(r)}$ and $\varphi_{2(r)}$ is given by $\theta$ as a function of coupling strength and detuning. The detuning can be caused by unintentional disorder in the fabricated structure such as random shifts in the position of the holes inevitably resulting in non-identically defined cavities. For nearly identical cavities, that is $\Delta\lambda \sim 0$ we obtain $\Delta\Omega \sim 2J$ from Eq. (1). Hence in this case the splitting is a direct measure of the coupling strength in the PM. For $\Delta\lambda \sim 0$ Eq. (2) tells us that the PM supports two supermodes resulting from equal contributions of the modes in isolated cavities as was shown numerically in Figs. 1(b) and 1(c). This corresponds to the strongly coupled device demonstrated in Fig. 3(c) with splitting 1.3 nm and for which a coupling strength $J \sim$ 0.65 nm (122 GHz) is deduced. Note that we found the effect of loss splitting [18] on the calculated $J$ to be negligible according to the measured Q for the S and AS supermodes shown in Fig. 3(c), 6300 and 5500, respectively. This coupling strength is comparable with results reported elsewhere for in-line coupled L3 [18] and heterostructure PhC cavities [15]. Here the intensity distribution gives us direct evidence of a near perfectly coupled system. Note that the symmetry of $\psi_{+(r)}$ and $\psi_{-(r)}$ are not known a priori. However, Figs. 3(c) and 3(d) show that for our particular PM with 6 lattice units centre-to-centre distance between the cavities, the S mode is the ground mode; this is important to specify as unlike atomic molecules the ground mode of PhC molecules can be AS depending on the distance between the cavities [31,32]. For example an AS ground mode was found in the case of in-line coupled L3 cavities for the same distance centre-to-centre [18,32]. For very dissimilar cavities we have $\Delta\lambda \gg 2J$, hence $\Delta\Omega \sim \Delta\lambda$. For such large detuning, the splitting is largely due to structural disorder. Also from Eq. (2) we get $\Psi_{+}(r) = \varphi_1(r)$ and $\Psi_{-}(r) = \varphi_2(r)$ i.e. the mode distribution of a weakly coupled system resembles that of two isolated cavities. This is observed experimentally in Fig. 4(d) for a splitting $\Delta\Omega = 2.8$ nm. Assuming $J \sim 0.65$ nm as obtained from the strongly coupled system, we deduce $\Delta\lambda \sim 2.5$ nm from Eq. (1), hence confirming the large contribution from the detuning to the measured splitting for this weakly coupled system. Note that this is only an approximation of the true detuning as $J$ is not expected to be conserved. Indeed, structural disorder could also affect the tunneling barrier between the optical wells defining the PM. Overall the coupled modes showed great variations in their intensity distribution for relatively small splitting as observed when comparing devices with a splitting 1.8 nm and 2.2 nm ($\Delta\lambda <$ 2 nm) shown in Fig 4(b) and Fig. 4(c), respectively. This indicates that the mode profile might be particularly sensitive to small detuning as discussed below.

The effect of deviations in the nominal dimensions of the cavities on the splitting and supermode profile are studied here with 3D FDTD and compared with a semi-analytical approach considering coupled oscillators. Extensive convergence tests were performed in order to resolve a ~1 nm change in the hole position. It was possible to resolve such a change by using a resolution of 30 pixels per lattice unit, provided a subpixel smoothing feature was employed during computation [23,33]. In the PM considered, the PhCWG modulation of the left cavity is varied as schematically shown in Fig. 5(a). First the detuning $\Delta\lambda = \lambda_1 - \lambda_2$ is obtained from isolated cavities where $\lambda_1$ and $\lambda_2$ are the wavelength of the right and left cavity resonances, respectively, computed with FDTD. A linear relation is obtained between structural variation and detuning as shown in Fig. 5(b), with a shift in the hole position by 1 nm resulting in $|\Delta\lambda| \sim 1$ nm. The resonant wavelengths of the coupled cavity modes are then computed to produce the splitting curve shown in Fig. 5(c). These results are compared with the data produced from Eq. (1) with the wavelength resonances computed from isolated cavities and a coupling strength $J = 0.9$ nm obtained from the coupled cavity system at zero detuning.

For the detuning range studied in this work we found ultra-high $Q > 10^7$ due to extremely low out-of-plane losses for coupled and detuned cavities, irrespective of the supermode symmetry. We also found $Q > 10^7$ for single cavities for all modulations of the waveguide width studied in this work, in good agreement with the results reported in [25] by considering cavities separated by 12 lattice units from the edge of the PhC. The effect of loss splitting and loss parameters mentioned in [18] were also found to be negligible on the calculated splitting, therefore losses were neglected in this model. We believe that the much lower Q measured experimentally for all devices presented in this work (between 5000 and 7000) to be the result of absorption by the quantum dot ensemble [27] as well as light scattering mechanisms such as the surface roughness resulting from the dry-etch process, not taken into account in the numerical model. Overall both calculations agree reasonably well over the detuning range studied although the semi-analytical model based on coupled-oscillators with first order perturbation [0] slightly overestimates the splitting with detuning. It is important to specify that the simple model based on coupled-cavity oscillators presented here assumes that the mode distribution of an isolated cavity can be approximated by the spatial distribution of the mode in each cavity of the coupled system. For the closely spaced cavities studied in this work this results in underestimated mode overlap between the resonant modes. It is especially apparent for zero detuning where the semi-analytical model results in smaller constructive and destructive interferences in the tunnelling barrier for the S and AS supermodes, respectively, as was seen in Fig. 1(b) and to a lesser extent in Fig. 1(c). We believe this discrepancy to be reduced for cavities farther apart. Overall these results show that the detuning largely contributes to the splitting for $|\Delta\lambda| > 2$ nm as found experimentally.

The time averaged electric field energy distribution of the coupled-cavity system together with the mode profile obtained from the convolution process are shown for increasing detuning in Figs. 6(a)-6(d). This is again compared with a semi-analytical approach using Eq. (2) to extract the weighting factors with results shown in Figs. 6(e)-6(h). For this method, the mode distribution is obtained from the FDTD computation of isolated cavities. Results shown here correspond to $\Delta\lambda < 0$ hence an increase in the PhCWG modulation width of the left cavity in which the low energy mode localizes with detuning. Results for $\Delta\lambda > 0$ are very similar and can be simply obtained by mirror reflection about the *y*-axis. Both methods predict strong mode localization for $|\Delta\lambda| > 2$ nm and overall show qualitatively similar mode distribution. The stronger mode localization with detuning given by the semi-analytical model is expected from the underestimated mode overlap. The mode delocalization can be expressed as the ratio of the electric field energy found at the centre of one cavity to that of the other with greater field energy, which we will refer to as the delocalization factor. It is clear that for very small detuning this factor is near unity and asymptotically approaches zero for large detuning, indicating a nearly fully localized mode in one of the cavities. Using Eq. (2) and neglecting the cross-terms by assuming relatively well confined modes in isolated cavities, the delocalization factor is given by:

$$\Gamma = \frac{\left(\sin\dfrac{\theta}{2}\right)^2}{\left(\cos\dfrac{\theta}{2}\right)^2} \tag{3}$$

and hence purely depends on the ratio $2J/\Delta\lambda$ given in Eq. (2). The results are plotted in Fig. 7 as a function of detuning for $J = 0.9$ nm together with the delocalization factor given by full FDTD calculations. We note that the full numerical computation predicts stronger localization with detuning for the low energy mode but a weaker one for the higher energy mode. Such differences are not predicted by the semi-analytical approach even when the cross-terms are taken into account. The following explanation is proposed: we recall here that the detuning is obtained by varying the modulation width of one of the cavities, that is the depth of the optical well as shown schematically in the inset of Fig. 7. Also, we know that the low energy mode localizes in the cavity with the deepest optical well. Therefore as the detuning increases, its confinement increases reducing the evanescent part of the mode found in the other cavity. On the other hand this picture is reversed for the high energy mode. Again this effect is expected to be less pronounced for cavities farther apart. However, it is clear that both models predict a very strong increase in mode localization for detuning up to 1 nm followed by a more gentle increase for $|\Delta\lambda| > 1$ nm as was seen in Figs. 4(b)-(d). We also note that the high energy modes retain some of the characteristics of the AS supermode even for relatively large detuning, i.e. a maxima in the intensity distribution in each cavity in accordance with the simulation results given in Fig. 6 and as seen experimentally in Figs. 4(b) and 4(c).

Hence we can conclude that the PM studied here transitions from a strongly coupled system for $\Delta\lambda < 1$ nm to a weakly coupled one for $\Delta\lambda > 2$ nm. For $\Delta\lambda < 1$ nm we obtain $\Delta\lambda < J$ as expected in the strong coupling regime. This transition can be the result of a shift in the hole position defining the modulation width of the cavity by 1 nm to 2 nm. We should add that the semi-analytical approach presented in this work resulted in significantly less computational time compared with the full FDTD computation of the detuned coupled cavities due to the preservation of symmetry and reduced cell size when computing the resonant modes of isolated cavities. Also, the small splitting between the supermodes greatly contributes to the time-consuming computation required to discriminate between these modes when calculations are performed entirely on disordered coupled cavities.

In summary we have demonstrated clearly two strongly coupled in-line cavities defined by small perturbations of a PhC waveguide by mapping the intensity distribution of the S and AS supermodes in real space using PL confocal microscopy. This technique extends previous transmission measurements on these waveguide cavities by demonstrating coupled cavities based on the field distribution. We also complement work done on PhC molecules of a different nature [21] by studying the effect of detuning on the mode delocalization both experimentally and theoretically, which we were able to quantify using FDTD and a semi-analytical approach. In particular, the clear distinction between the S and AS supermodes in the PM presented here enabled us to investigate the transition from a strongly to a weakly coupled system in the presence of detuning due to structural variations.

Such variations were simulated by minute deviations in the perturbation of the PhC waveguide using 3D FDTD. The semi-analytical method mainly consisted in computing the resonances of isolated cavities with FDTD and of a perfectly coupled system to extract the coupling strength. This approach was shown to be more time-efficient in predicting the strong variations seen in the mode localization with small detuning. This enabled us to show that the device preserves the characteristics of a strongly coupled system for detuning smaller than 1 nm, corresponding to about 1 nm shift in the hole position defining the cavity. Also, as the PL mapping is able to pinpoint the centre of each cavity via the distribution of the resonant modes, a post-processing control of the detuning could be employed to systematically obtain strongly coupled cavities [34]. Finally we would like to emphasize that the theoretical $Q > 10^7$ for all coupled or detuned cavities studied in this work makes this system particularly appealing for strong light matter coupling experiments. We believe that the results presented in this study are particularly relevant for applications in cQED where the local strength of the electric field dictates the coupling to a quantum emitter.

**Acknowledgements**

*This work was performed using the Darwin Supercomputer of the University of Cambridge High Performance Computing Service (http://www.hpc.cam.ac.uk/), provided by Dell Inc. using Strategic Research Infrastructure Funding from the Higher Education Funding Council for England*

*and funding from the Science and Technology Facilities Council. The authors acknowledge support from EPSRC on grant EP/K012029/1 "Long wavelength single photon sources and dotonic molecules". The authors would like to thank Dr. Dimitris Angelakis and Dr. Andrew Ramsey for invaluable discussions and support.*

**References**

[1]. M. Bayer, T. Gutbrod, J. P. Reithmaier, A. Forchel, T. L. Reinecke, P. A. Knipp, A. A. Dremin, and V. D. Kulakovskii, "Optical modes in photonic molecules," Phys. Rev. Lett. **81**, 2582-2585 (1998).

[2]. S. Ishii and T. Baba, "Bistable lasing in twin microdisk photonic molecules," Appl. Phys. Lett. **87**, 181102-1-3 (2005).

[3]. S. V. Zhukovsky, D. N. Chigrin, A. V. Lavrinenko, and J. Kroha, "Switchable lasing in multimode microcavities," Phys. Rev. Lett. **99**, 073902-1-4 (2007).

[4]. S. Fan, P. R. Villeneuve, J. D. Joannopoulos, and H. A. Haus, "Channel drop filters in photonic crystals," Opt. Express **3**, 4–11 (1998).

[5]. E. I. Simakov, L. M. Earley, C. E. Heath, D. Yu. Shchegolkov, and B. D. Schultz, "First experimental demonstration of a photonic band gap channel-drop filter at 240 GHz," Rev. Sci. Inst. **81**, 104701-1-5 (2010).

[6]. M. T. Hill, H. J. S. Dorren, T. de Vries, X. J. M. Leijtens, J. H. den Besten, B. Smalbrugge, Y. S. Oei, H. Binsma, G. D. Khoe, and M. K. Smit , "A fast low-power optical memory based on coupled micro-ring lasers," Nature **432**, 206-208 (2004).

[7]. A. Dousse, J. Suffczynski, A. Beveratos, O. Krebs, A. Lemaitre, I. Sagnes, J. Bloch, P. Voisin, and P. Senellart, "Ultrabright source of entangled photon pairs," Nature **466**, 217–220 (2010).

[8]. T. C. H. Liew and V. Savona, "Single photons from coupled quantum modes," Phys. Rev. Lett. **104**, 183601-1-4 (2010).

[9]. M. Bamba, A. Imamoğlu, I. Carusotto, and C. Ciuti, "Origin of strong photon antibunching in weakly nonlinear photonic molecules," Phys. Rev. A **83**, 021802(R)-1-4 (2011).

[10]. M. Bamba and C. Ciuti, "Counter-polarized single-photon generation from the auxiliary cavity of a weakly nonlinear photonic molecule," Appl. Phys. Lett. **99**, 171111-1-3 (2011).

[11]. Y. Akahane, T. Asano, B-S. Song, and S. Noda, "High-Q photonic nanocavity in a two-dimensional photonic crystal," Nature **425**, 944-947 (2003).

[12]. M. Notomi, E. Kuramochi, and T. Tanabe, "Large-scale arrays of ultrahigh-Q coupled nanocavities," Nat. Photonics **2**, 741-747 (2008).

[13]. M. J. Hartmann, F. G. S. Brandão, and M. B. Plenio, "Strongly interacting polaritons in coupled arrays of cavities," Nat. Physics **2**, 849-855 (2006).

[14]. D. G. Angelakis, M. F. Santos, and S. Bose, "Photon-blockade-induced Mott transitions and XY spin models in coupled cavity arrays," Phys. Rev. A **76**, 031805-1-4 (2007).

[15]. D. O' Brien, M. D. Settle, T. Karle, A. Michaeli, M. Salib, and T. F. Krauss, "Coupled photonic crystal heterostructure nanocavities," Opt. Express **15**, 1228-1233 (2007).

[16]. S. Vignolini, F. Intonti, M. Zani, F. Riboli, D. S. Wiersma, L. H. Li, L. Balet, M. Francardi, A. Gerardino, A. Fiore, and M. Gurioli, "Near-field imaging of coupled photonic-crystal Microcavities," Appl. Phys. Lett. **94**, 151103-1-3 (2009).

[17]. S. Lam, A. R. Chalcraft, D. Szymanski, R. Oulton, B. D. Jones, D. Sanvitto, D. M. Whittaker, M. Fox, M. S., Skolnick, D. O'Brien, T. F. Krauss, H. Liu, P. W. Fry, and M. Hopkinson, "Coupled resonant modes of dual L3-defect planar photonic crystal cavities," in *Conference on Lasers and Electro-Optics/Quantum Electronics and Laser Science Conference and Photonic Applications Systems Technologies, OSA Technical* Digest (CD) (Optical Society of America, 2008), paper QFG6. http://www.opticsinfobase.org/abstract.cfm?URI=QELS-2008-QFG6.

[18]. K. A. Atlasov, K. F. Karlsson, A. Rudra, B. Dwir, and E. Kapon, "Wavelength and loss splitting in directly coupled photonic-crystal defect microcavities," Opt. Express **16**, 16255-16264 (2008).

[19]. M. Brunstein, T. J. Karle, I. Sagnes, F. Raineri, J. Bloch, Y. Halioua, G. Beaudoin, L. Le Gratiet, J. A. Levenson, and A. M. Yacomotti, "Radiation patterns from coupled photonic crystal nanocavities ," Appl. Phys. Lett. **99**, 111101-1-3 (2011).

[20]. A. Majumdar, A. Rundquist, M. Bajcsy, and J. Vuckovic, "Cavity quantum electrodynamics with a single quantum dot coupled to a photonic molecule," Phys. Rev. B **86**, 045315-1-5 (2012).

[21]. K. Foubert, B. Cluzel, L. Lalouat, E. Picard, D. Peyrade, F. de Fornel, and E. Hadji, "Influence of dimensional fluctuations on the optical coupling between nanobeam twin cavities," Phys. Rev. B **85**, 235454-1-7 (2012).

[22]. L. C. Andreani, D. Gerace, and M. Agio, "Exciton-polaritons and nanoscale cavities in photonic crystal slabs," Phys. Status Solidi B **242**, 2197-2209 (2005).

[23]. A. F. Oskooi, D. Roundy, M. Ibanescu, P. Bermel, J. D. Joannopoulos, and S. G. Johnson, "MEEP: A flexible free-software package for electromagnetic simulations by the FDTD method," Comput. Phys. Commun. **181**, 687–702 (2010).

[24]. V. A. Mandelshtam and H. S. Taylor, "Harmonic inversion of time signals," J. Chem. Phys. **107**, 6756-6769 (1997). Erratum, ibid. **109**, 4128 (1998).

[25]. E. Kuramochi, M. Notomi, M. Mitsugi, A. Shinya, T. Tanabe, and T. Watanabe, "Ultrahigh-Q photonic crystal nanocavities realized by the local width modulation of a line defect," Appl. Phys. Lett. **88**, 041112-1-3 (2006).

[26]. B. S. Song, S. Noda, T. Asano, and Y. Akahane, "Ultra-high-Q photonic double-heterostructure nanocavity," Nat. Mater. **4**, 207-210 (2005).

[27]. F. S. F. Brossard, X. L. Xu, D. A. Williams, M. Hadjipanayi, M. Hugues, M. Hopkinson, X. Wang, and R. A. Taylor, "Strongly coupled single quantum dot in a photonic crystal waveguide cavity," Appl. Phys. Lett. **97**, 111101-1-3 (2010).

[28]. K. H. Lee, A. M. Green, R. A. Taylor, D. N. Sharp, J. Scrimgeour, O. M. Roche, J. H. Na, A. F. Jarjour, A. J. Turberfield, F. S. F. Brossard, D. A. Williams, and G. A. D. Briggs, "Registration

of single quantum dots using cryogenic laser photolithography," Appl. Phys. Lett. **88**, 193106-1-3 (2006).

[29]  D. C. Reynolds, K. K. Bajaj, C. W. Litton, G. Peters, P. W. Yu, and J. D. Parsons, "Refractive index, n, and dispersion, −dn/dλ, of GaAs at 2 K determined from Fabry–Perot cavity oscillations," J. Appl. Phys. **61**, 342-345(1987).

[30].  C. Cohen-Tannoudji, B. Diu, and F. Lalöe, *Quantum Mechanics* (Wiley-Interscience, Paris, 1977), Chap. 4.

[31].  N. Caselli, F. Intonti, F. Riboli, A. Vinattieri, D. Gerace, L. Balet, L. Li, M. Francardi, A. Gerardino, A. Fiore, and M. Gurioli, "Antibonding ground state in photonic crystal molecules," Phys. Rev. B **86**, 035133-1-4 (2012).

[32]  A. R. A. Chalcraft, S. Lam, B. D. Jones, D. Szymanski, R. Oulton, A. C. T. Thijssen, M. S. Skolnick, D. M. Whittaker, T. F. Krauss, and A. M. Fox, "Mode structure of coupled L3 photonic crystal cavities," Opt. Express **19**, 5670-5675 (2011).

[33]  C. Kottke, A. Farjadpour, and S. G. Johnson, "Perturbation theory for anisotropic dielectric interfaces, and application to subpixel smoothing of discretized numerical methods," Phys. Rev. E **77**, 036611-1-10 (2008).

[34]  N. Caselli, F. Intonti, C. Bianchi, F. Riboli, S. Vignolini, L. Balet, L. H. Li, M. Francardi, A. Gerardino, A. Fiore, and M. Gurioli, "Post-fabrication control of evanescent tunnelling in photonic crystal molecules," Appl. Phys. Lett. **101**, 211108-1-5 (2012).

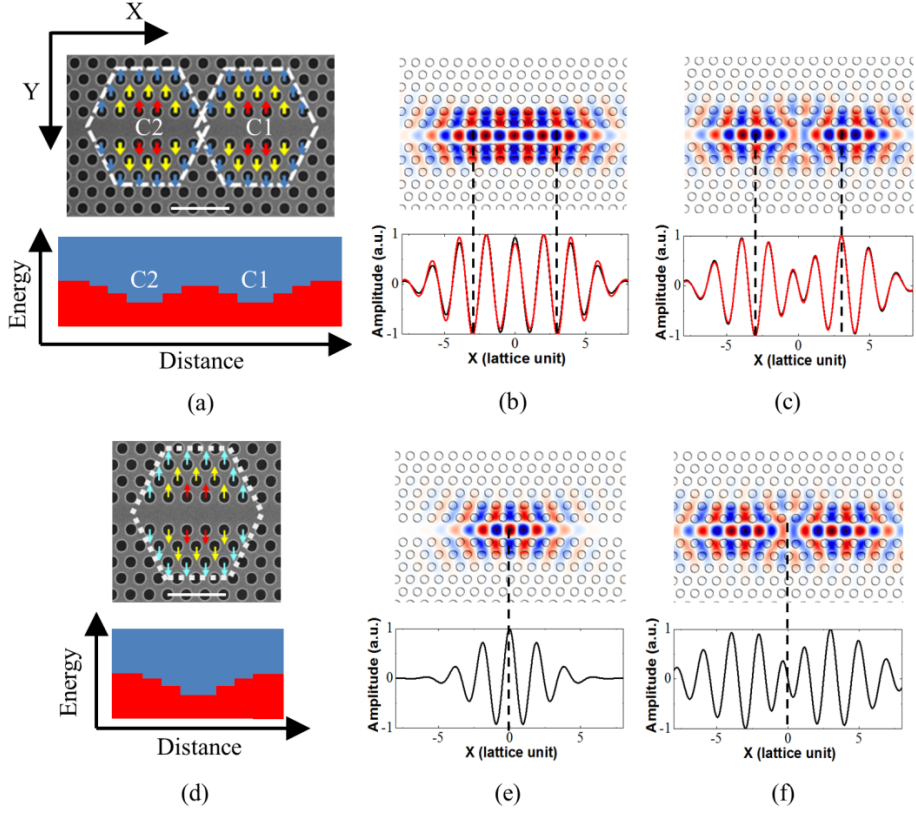

FIG. 1. (Color) Calculated mode profiles for two-coupled and a single PhCWG cavity defined by the local modulation of the waveguide width. The waveguide width outside the cavity is given by $W = 0.98\sqrt{3}a$ where $a$ is the lattice constant. (a) Top: SEM image of the PM. Scale bar: 1 μm. The holes are shifted within each dashed hexagon by (red arrows) $dy = 12$ nm, (yellow arrows) 8 nm ($2dy/3$), and (blue arrows) 4 nm ($dy/3$) in a 330 nm PhC lattice with hole size ~ 180 nm. Bottom: Schematic band diagram of the PM showing the stop gap in red. (b-c) $Ey$ Electric field distribution of the supermodes and their profile along the waveguide direction. The dashed lines indicate the centre of each cavity. The S mode is shown in (b) and AS mode in (c). The profile in black is obtained from 3D FDTD calculations on the PM, that in red superimposed from φ(x)+φ(x-6a) for the S mode and φ(x)-φ(x-6a) for the AS mode where φ(x) corresponds to the 3D FDTD mode profile of a single cavity and 6a the centre-to-centre distance between the cavities. (d) Same as (a) but for a single cavity. (e) $Ey$ Electric field distribution of the first order mode of a single cavity and its profile φ(x) along the waveguide direction. (f) Same as (e) but for the second order mode of the single cavity.

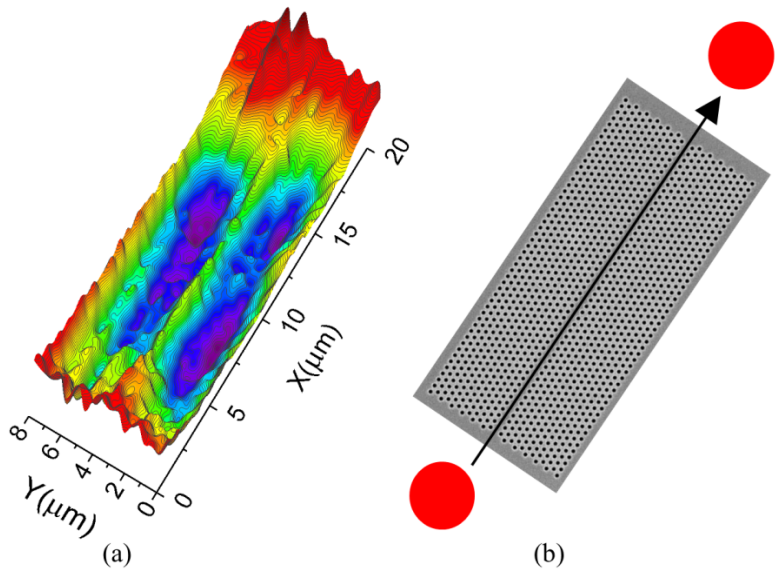

(a)                                             (b)

FIG. 2. (Color) (a) Integrated PL mapping of a PhCWG without cavity. (b) SEM image of the PhCWG. The PL mapping is obtained by scanning the laser spot pumping the QDs along the direction of the waveguide defined by the *X* axis then translated along the *Y* axis.

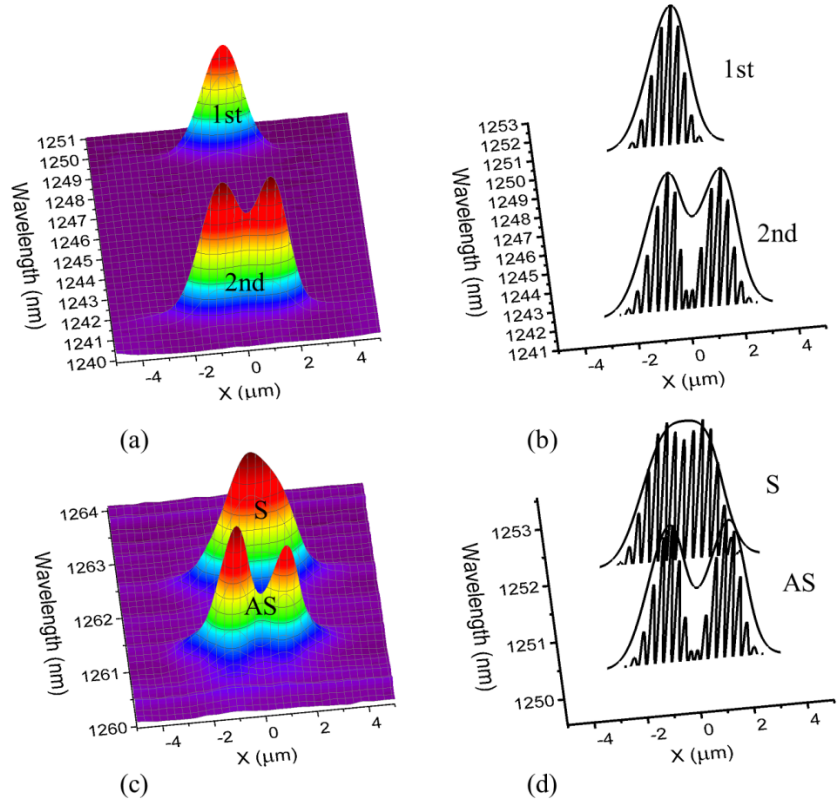

FIG. 3. (Color) Experimental demonstration of a strongly coupled system in a PM involving the first order mode of a PhCWG cavity. (a) PL spectra of a single cavity collected along the waveguide showing the first and second order bound mode of the cavity. (b) Predicted mode profile for the first and second order bound mode of a single cavity as obtained from 3D FDTD calculations. The time averaged electric field energy distribution is shown in black. The convoluted profile with a Gaussian beam of 1.2 μm size is shown by the envelope. (c) PL spectra of two-coupled cavities collected along the waveguide showing the first order S and AS supermodes. (d) Predicted time averaged electric field energy distribution and convoluted mode profile for the first order S and AS supermodes of two-coupled cavities as obtained from 3D FDTD calculations.

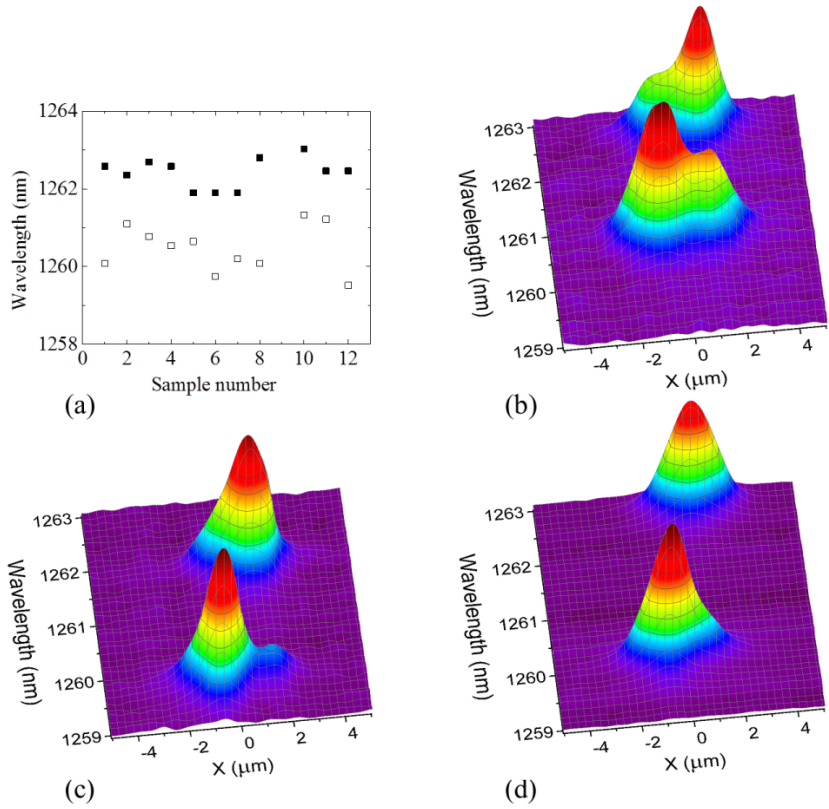

FIG. 4. (Color) Effect of random structural variations in fabricated PMs on the profile of the supermodes of two-coupled PhCWG cavities. (a) High and low energy supermodes of a batch of nominally identical coupled cavity systems measured from the PL spectra. (b-d) PL spectra of coupled cavities obtained along the waveguide direction for each device and presented in increasing order of splitting $\Delta\Omega$ = 1.8 nm, 2.2 nm and 2.8 nm, respectively.

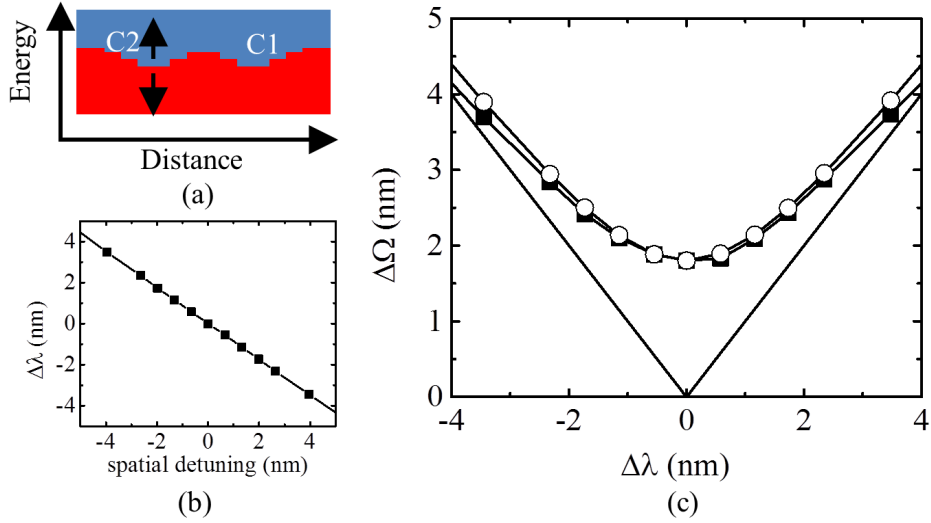

(a)

(b)

(c)

FIG. 5. (Color) Calculated splitting as a function of detuning for two-coupled PhCWG cavities. (a) Schematic band diagram showing the variation in the modulation width of the left cavity C2. In this process the holes indicated by the red arrows in Fig. 1 are spatially shifted by a few nm in the $y$ direction. The other holes of the left cavity are shifted accordingly as to maintain the ratio described in Fig. 1.(b) Calculated detuning with 3D FDTD as a function of the variation in the modulation width (spatial detuning) of the left cavity. (c) Calculated splitting $\Delta\Omega$ as a function of detuning $\Delta\lambda$. Results from the full 3D FDTD computation on the detuned system (plain squares) are compared with that obtained from a semi-analytic approach (circles). The straight lines correspond to the difference in the resonant wavelength between isolated (or uncoupled) cavities, hence $\Delta\Omega = \Delta\lambda$ where one of them experiences the same variation in the modulation width as C2.

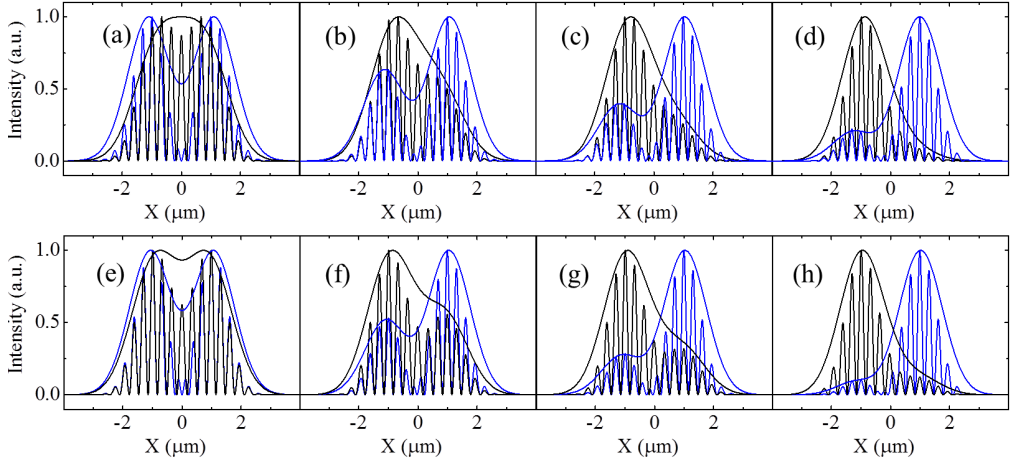

FIG. 6. (Color) Evolution of the calculated mode profile of the S (black) and AS (blue) supermodes as a function of detuning for two-coupled PhCWG cavities showing the time averaged electric field energy distribution and the envelope resulting from the convolution with a Gaussian beam of spot size 1.2 μm. The detuning from left to right corresponds to $\Delta\lambda$ = 0 nm, 0.6 nm, 1.2 nm, and 2.3 nm, respectively. (a-d) Results obtained from full 3D FDTD computation on the detuned system. (e-h) Results obtained from the semi-analytic approach.

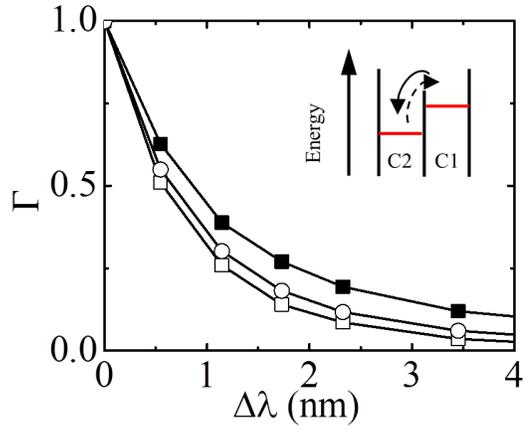

FIG. 7. (Color) Calculated supermode delocalization factor $\Gamma$ as a function of detuning $\Delta\lambda$ for two-coupled PhCWG cavities. Results from the semi-analytical approach (circles) are compared with results from the full 3D FDTD on the detuned system for the high (plain squares) and low (empty squares) energy modes. A mechanism to explain the difference between the mode delocalization of the high and low energy modes with detuning is shown schematically in the inset.